\documentclass[dvips]{PoS}
\usepackage{subfigure}
\usepackage{amsmath}

\newcommand{\tr}{\operatorname{Tr}}
\newcommand{\re}{\operatorname{Re}}

\title{First results with two light flavours of quarks with maximally twisted mass}

\ShortTitle{First results with $N_f=2$ maximally twisted mass quarks }

\author{Karl Jansen$^*$\\
        John von Neumann Institute for Computing, NIC, \\
        Platanenallee 6, D-15738 Zeuthen, Germany\\
        E-mail: \email{Karl.Jansen@desy.de}}

\author{\speaker{Carsten Urbach}\\
        Theoretical Physics Division, Dept. of Mathematical Sciences,\\
        University of Liverpool, Liverpool, L69 7ZL, U.K.\\
        E-mail: \email{Carsten.Urbach@Liverpool.ac.uk}}

\author{On behalf of the European Twisted Mass Collaboration (ETMC)}

\abstract{We report on first results of an ongoing effort to simulate
  lattice QCD with two degenerate flavours of quarks by means of the
  twisted mass formulation tuned to maximal twist. 
  By utilising recent improvements of the
  HMC algorithm, pseudo-scalar masses well below $300\, \mathrm{MeV}$
  are simulated on volumes $L^3\cdot T$ with $T=2L$ and 
  $L> 2\, \mathrm{fm}$ and at 
  values of the lattice spacing $a\lesssim 0.1\, \mathrm{fm}$. 
  We present first evidence that scaling
  violations in the pseudo-scalar decay constant are small and well compatible
  with $\mathcal{O}(a)$ improvement. Additionally, exploratory results for the case 
  of $N_f=2+1+1$ flavours are discussed.}

\FullConference{XXIVth International Symposium on Lattice Field Theory\\
  23-28 July 2006\\
  Tucson, Arizona, USA}

\begin{document}

\section{Introduction}

Originally, the twisted mass formulation of lattice QCD
\cite{Frezzotti:1999vv,Frezzotti:2000nk} was invented to
provide the Wilson lattice Dirac operator with an infra-red cut-off for its
eigenvalue spectrum in order to prevent the appearance of unphysically
small eigenvalues when the quark masses become small. 
More recently, however, another quite astonishing property of twisted
mass quarks has been  discovered: in Ref.~\cite{Frezzotti:2003ni} it
was shown that pure Wilson twisted mass fermions lead to 
\emph{automatic $\mathcal{O}(a)$ improvement} when tuned to full twist
\emph{even without any improvement term added to the action.}      

In simulations in the quenched approximation it has been demonstrated
that the presence of the twisted mass term as an infra-red regulator
is indeed a clear advantage when compared to the pure Wilson
regularisation \cite{Jansen:2003ir}. However, this infra-red regulator
appears, somewhat surprisingly, not to be crucial in simulations with
light dynamical quarks. Here simulations even with standard Wilson
fermions at pseudo-scalar masses below $300\, \mathrm{MeV}$ are
possible, at least when newer algorithmic techniques are 
used, see e.g. our own work of Ref.~\cite{Urbach:2005ji}.

It appears therefore that for the case of \emph{dynamical} simulations
the aforementioned automatic $\mathcal{O}(a)$ improvement of the  
(maximally twisted) Wilson formulation of lattice QCD becomes more 
important\footnote{Naturally, having an infra-red cut-off in the theory does 
guarantee a stable simulation, also in difficult situations. We consider therefore 
the presence of the twisted mass term still to be an advantage. In particular, 
when we think, e.g.,  of simulations much closer to the physical point than performed 
today, the presence of the twisted mass term may turn out to be crucial.}.
In  such a set-up only one parameter needs to be tuned in order to
obtain $\mathcal{O}(a)$ improvement and there are no additional
operator-specific improvement terms needed. Moreover, mixing patterns
during renormalisation are greatly simplified. 

The disadvantage of the twisted mass formulation is the fact
that the twisted mass term explicitly breaks parity and flavour
symmetry. The flavour symmetry breaking for instance introduces a
mass splitting between the lightest charged and the uncharged
pseudo-scalar mesons. However, all those effects stem only from the
valence quark sector \cite{Frezzotti:2003xj} (the sea quark
determinant is flavour blind), are  
expected to be lattice
artifacts of order $\mathcal{O}(a^2)$. In particular, in a mixed
action set-up, with for instance overlap valence fermions no breaking
of parity or flavour chiral symmetry is expected.

In the quenched approximation it is by now established that
automatic $\mathcal{O}(a)$ improvement works very well in practice
\cite{Jansen:2003ir,Jansen:2005gf,Jansen:2005kk,Abdel-Rehim:2004gx,Abdel-Rehim:2005gz} .
Also, the theoretical 
expectation of $\mathcal{O}(a^2)$ lattice artifacts appearing in the 
pseudo-scalar mass splitting has been confirmed 
\cite{Jansen:2005cg,Farchioni:2005hf}. 
In this proceeding contribution we will
provide first evidence that maximally twisted mass quarks are
$\mathcal{O}(a)$ improved also in dynamical simulations and that the
residual $\mathcal{O}(a^2)$ lattice artifacts are small. The lowest
value of the pseudo-scalar mass covered in these simulations is
significantly below $300\, \mathrm{MeV}$, which is in a region 
where it is possible to confront the simulation data with chiral
perturbation theory ($\chi$PT) 
and determine therefore the validity range of $\chi$PT. 
Since simulations directly at the physical value of the pion mass are still 
somewhat too expensive, $\chi$PT could then be used
to bridge the gap between the simulation data and 
the physical point. 
The low values of the pseudo-scalar
mass reached in our work became possible only due to recent algorithmic developments
\cite{Hasenbusch:2001ne,Hasenbusch:2002ai,Peardon:2002wb,Luscher:2004rx,Urbach:2005ji,Jansen:2005yp,Clark:2006fx,AliKhan:2003br}
(See also the contribution by M.~Clark to these proceedings).
In particular, for all the $N_f=2$ results reported in this proceeding contribution
we have used our variant of the HMC algorithm described in Ref.~\cite{Urbach:2005ji}.

In addition to observables like pseudo-scalar mass and decay constant 
we will provide also first results for the
aforementioned mass splitting in the pseudo-scalar sector and discuss 
the case of $N_f=2+1+1$ flavours of quarks, i.e. taking the strange
and the charm quark into account in the simulation.
For recent summaries of the status of Wilson twisted mass fermions see
Refs.~\cite{Farchioni:2005ec,Scorzato:2005rb}, the review of 
Ref.~\cite{Shindler:2005vj} and references therein.

\section{Choice of Lattice Action}

The fermionic lattice action for two flavours of mass degenerate
Wilson twisted mass quarks reads (in the so called twisted basis $\chi$)
\begin{equation}
  \label{eq:Sf}
  \begin{split}
    S_\mathrm{tm} =\  a^4\sum_x\biggl\{ &
      \bar\chi_x[am_0+4r + i\gamma_5\tau_3a\mu]\chi_x \Bigr. \\
      & \Bigl. +\sum_{\nu=1}^4
      \bar\chi_x\left[U_{x,\nu}(r+\gamma_\nu)\chi_{x+\hat\nu} +
        U_{x-\hat\nu,\nu}^\dagger(r-\gamma_\nu) \chi_{x-\hat\nu}\right]\biggr\}\, ,
  \end{split}
\end{equation}
where $am_0$ is the bare untwisted quark mass and $a\mu$ the bare twisted
mass. $\tau_3$ is the third Pauli matrix acting in flavour space 
and $r$ is the Wilson parameter, which we set to one in our
simulations. 

Twisted mass fermions are said to be at maximal twist if the bare
untwisted mass is tuned to its critical value $m_\mathrm{crit}$. We
will discuss later on how this can be realised in practice.

In the gauge sector we use the so called tree-level Symanzik improved
gauge action (tlSym) \cite{Weisz:1982zw}, which includes besides the
plaquette term $U_\Box$ also rectangular $(1\times2)$ Wilson loops
$U^{1\times2}_{x,\mu,\nu}$
\begin{equation}
  \label{eq:Sg}
    S_g =  \beta\sum_x\Biggl(  b_0\sum_{1\leq\mu<\nu;
      \mu,\nu=1}^4\frac{1}{3}\{1-\re\tr(U_\Box)\}\Bigr. 
     \Bigl.\ +\ 
    b_1\sum_{\mu\neq\nu;\mu,\nu=1}^4\frac{1}{3}\{1
    -\re\tr(U^{1\times2}_{x,\mu,\nu})\}\Biggr)\, , 
\end{equation}
with the bare inverse coupling $\beta$, $b_1=-1/12$ and the
normalisation condition $b_0=1-8b_1$. Note that with $b_1=0$ this
action corresponds to the usual Wilson plaquette action.

Before presenting the actual results let us first briefly recall 
some earlier works which led to the present production set-up, see also
Ref.~\cite{Farchioni:2005ec}.
As mentioned before, in
the quenched approximation we have shown in an extended scaling test
that $\mathcal{O}(a)$ improvement works extremely well for maximally
twisted mass quarks
\cite{Jansen:2003ir,Jansen:2005gf,Jansen:2005kk}. In the context of
this scaling test also several definitions for full twist were
examined and it was found that in agreement with theoretical
considerations \cite{Frezzotti:2005gi,Aoki:2004ta,Sharpe:2004ny} the
so called PCAC definition 
leads to small residual lattice artifacts of $\mathcal{O}(a^2)$ in the
whole range of -- in particular, also small --  masses investigated: 
for the PCAC definition, at a fixed value of $a\mu > 0$, 
$am_\mathrm{crit}$ is determined from the condition that                     
\begin{equation}
  \label{eq:mpcac}
  m_\mathrm{PCAC} = \frac{\sum_\mathbf{x}\langle\partial_0 A_0^a(x)
    P^a(0)\rangle} {2\sum_\mathbf{x}\langle P^a(x) P^a(0)\rangle}\qquad
  a=1,2
\end{equation}
vanishes\footnote{Note that $m_\mathrm{PCAC}$ is sometimes
  denoted with $m_\chi^\mathrm{PCAC}$, because the bilinears are defined
  in the twisted basis $\chi$.} for large enough time separation such that
a projection on the pion state is reached. This provides then $am_\mathrm{crit}$
as a function of $a\mu$ and the final value of
$am_\mathrm{crit}$ may be obtained by (linearly) 
extrapolating $am_\mathrm{crit}(\mu)$ to $\mu=0$. 
However, the extrapolation of $am_\mathrm{crit}(\mu)$ to $\mu=0$ is not
necessary: taking the value of $am_\mathrm{crit}$ at the lowest value
of $a\mu$ that is used in the simulation 
is an equally good definition. This remark is of particular
importance for dynamical simulations where one might not be able to
afford to determine $am_\mathrm{crit}$ at several values of 
$a\mu$ and then perform the extrapolation to $\mu=0$.

Furthermore, in the dynamical case it is not clear at all how an 
extrapolation to $a\mu=0$ could be performed: a generic property of
Wilson type fermions is its non-trivial phase structure at finite
lattice spacing with first order phase transitions 
close to the chiral limit, which was investigated in
detail by means of twisted mass fermions in
Refs.~\cite{Farchioni:2004us,Farchioni:2005tu,Farchioni:2004fs,Farchioni:2005ec} 
and which is 
in accordance with the picture developed in $\chi$PT 
\cite{Sharpe:1998xm,Munster:2004am}. 
A very important consequence of this
phenomenon is that at a non-vanishing lattice spacing the value of
the pseudo-scalar mass cannot drop below a certain minimal
value without being affected by the first order phase
transition. From the analysis in $\chi$PT it is expected that the
corresponding minimal value of the twisted mass parameter 
goes to zero proportional to $a^2$ towards the continuum limit. Hence, 
one can always find a largest value of the lattice spacing
$a_\mathrm{max}$ where simulations with a given value of the
pseudo-scalar mass can be performed.

The actual value of $a_\mathrm{max}$ depends on many details, 
for instance on
the gauge action used in the simulation. For example,
when the Wilson plaquette gauge action is used one can estimate 
$a_\mathrm{max}\approx 0.07-0.1\,\mathrm{fm}$ to realise a pseudo-scalar
mass of $250\,\mathrm{MeV}$.
The situation improves, 
when also the $2\times1$ rectangular loops are included
(see Eq.~(\ref{eq:Sg})): the value of $a_\mathrm{max}$ 
increases when the modulus of the coefficient $b_1$ multiplying the rectangular part
is increased \cite{Farchioni:2004fs,Farchioni:2005tu}. The study of
different gauge actions led us to the conclusion that the tlSym gauge
action, with a coefficient of $b_1=-1/12$, 
is a good choice to avoid on the one hand large effects
related to the presence of the phase transition and on the other hand the
possible conceptual issues concerning the DBW2 gauge action, with 
coefficient $b_1=-1.4088$, \cite{Farchioni:2005ec}.

\section{Set-up and First Results}

\begin{table}[t]
  \centering
  \begin{tabular*}{1.\linewidth}{@{\extracolsep{\fill}}lcccc}
    \hline\hline
    $\Bigl.\Bigr.\beta$ & $L^3\times T$ & $a\mu_\mathrm{min}$
    & $\kappa_\mathrm{crit}(a\mu_\mathrm{min})$& $r_0/a(a\mu_\mathrm{min})$ \\
    \hline\hline
    $3.9$  & $24^3\times48$ & $0.004$ & $0.160856$ & $5.184(41)$ \\
    $4.05$ & $32^3\times64$ & $0.003$ & $0.15701$  & $6.525(101)$ \\
    \hline\hline
  \end{tabular*}
  \caption{The parameters for the simulation. We denote by $a\mu_\mathrm{min}$ 
    the smallest value of the twisted mass parameter $a\mu$ at which we have performed
    simulations. At this value of $a\mu$ we determined the critical mass 
    $m_\mathrm{crit}$, or, equivalently
    the hopping parameter $\kappa_\mathrm{crit}=1/(8+2am_\mathrm{crit})$. 
    In order to convert to physical units, 
    we take the value of $r_0/a$ as computed at $a\mu_\mathrm{min}$.}
  \label{tab:setup}
\end{table}

The main goal of the present work is to investigate, whether also in the 
case of $N_f=2$ mass-degenerate flavours of dynamical quarks, the lattice 
artifacts are as small as has been found in the quenched approximation when 
maximally twisted Wilson fermions are used.
The current status of our simulations are summarised in tables
\ref{tab:setup} and \ref{tab:results}. 
While in table \ref{tab:setup} we provide the parameters of our simulations, 
in table \ref{tab:results} we also give 
current estimates of the pseudo-scalar mass $m_\mathrm{PS}$ in
physical units, where we used $r_0=0.5\,\mathrm{fm}$. In addition we
provide the number of equilibration trajectories $N_\mathrm{therm}$, 
the current statistics $N_\mathrm{traj}$ we have available and  
estimates of the integrated plaquette autocorrelation time where available.

The algorithm we used is a HMC algorithm with mass preconditioning
\cite{Hasenbusch:2001ne} and multiple time scale integration. It is
described in detail in Ref.~\cite{Urbach:2005ji}. The trajectory
length was set to $1/2$ in all our runs. Our currently available
estimates of the plaquette integrated autocorrelation time
$\tau_\mathrm{int}(P)$ quoted in table \ref{tab:results} are in units of
$\tau=1/2$.

\begin{table}[t]
  \centering
  \begin{tabular*}{1.\linewidth}{@{\extracolsep{\fill}}lccccc}
    \hline\hline
    $\Bigl.\Bigr.\beta$ & $a\mu$
    & $m_\mathrm{PS}\,[\mathrm{MeV}]$ & $N_\mathrm{therm}$ &
    $N_\mathrm{traj}$ & $\tau_\mathrm{int}(P)$\\
    \hline\hline
    $3.9$ & $0.0040$ & $280$ & $1500$ & $5000$ & $55(17)$\\

    & $0.0064$ & $350$ & $1500$ & $5000$ & $23(07)$\\

    & $0.0100$ & $430$ & $1500$ & $5000$ & $15(04)$\\

    & $0.0150$ & $530$ & $1500$ & $5000$ & $06(02)$\\
    \hline
    $4.05$ & $0.0030$ & $266$ & $1500$ & $2200$ & -\\

    & $0.0060$ & $372$ & $1500$ & $500$ & -\\
    \hline\hline
  \end{tabular*}
  \caption{Current status of the simulations at $\beta=3.9$ and $\beta=4.05$ using
    the simulation parameters listed in
    table~\protect\ref{tab:setup}. In addition we give the current estimates
    of $m_\mathrm{PS}$ in physical units ($r_0=0.5\,\mathrm{fm}$) and
    of $\tau_\mathrm{int}(P)$ where available.}
  \label{tab:results}
\end{table}

Full twist is realised in our simulations by tuning at a fixed value of
$\beta$ the PCAC quark mass $m_\mathrm{PCAC}$ to zero at the smallest
value $a\mu_\mathrm{min}$ of the twisted mass parameter $a\mu$. Keeping
then the value of $\kappa_\mathrm{crit}$ or, equivalently, 
$m_\mathrm{crit}$, for all the other values of
$a\mu$ provides $\mathcal{O}(a)$ improvement for all relevant physical
quantities (see
Refs.~\cite{Aoki:2004ta,Frezzotti:2005gi,Jansen:2005kk}). We list in table 
\ref{tab:setup} the values of $\kappa_\mathrm{crit}$ which we then kept fixed 
for all simulations at a given value of $\beta$. The dependence of 
$am_\mathrm{PCAC}$ on $a\mu$ is shown in figure \ref{fig:mpcac}.   
The fact that at values of $a\mu > a\mu_\mathrm{min}$ the PCAC quark mass 
$am_\mathrm{PCAC}$ does not vanish will only lead to effects that are
of $\mathcal{O}(a^2)$ and hence will not destroy $\mathcal{O}(a)$
improvement.

In order to make maximum use of the gauge configurations, we evaluate
connected meson correlators  using a stochastic method to include all
spatial sources. This method  involves a stochastic source (Z(2)-noise in both
real and imaginary part) for  all colours and spatial locations at one
Euclidean time slice. By solving for the propagator from this source  for 
each of the 4
spin components, we can construct zero-momentum meson correlators from
any quark bilinear at the source and sink. This method involves 4 inversions
per Euclidean time slice value selected since we chose to use only one 
stochastic sample.
This "one-end"  method is similar to that  pioneered in
Ref.~{\cite{Foster:1998vw} and implemented in Ref.~{\cite{McNeile:2006bz}}.
  We also employ a fuzzed source \cite{Lacock:1994qx} of extent 6 lattice
  spacings  to enable study of  non-local meson creation and destruction
  operators.

  In general, we save a gauge configuration every second trajectory
  and analyse meson correlators as described above from a selection of
  different Euclidean time slice sources. To reduce autocorrelations, we only use the
  same time slice source every 8-10 trajectories. Our primary statistical error
  analysis is from a bootstrap  analysis of blocks of
  configurations. This estimate was cross-checked by varying the block size.

  \begin{figure}[t]
    \centering
    \subfigure[$am_\mathrm{PCAC}$ as a function of $a\mu$ at
      $\beta=3.9$. \label{fig:mpcac}]
	      {\includegraphics[width=0.45\linewidth]{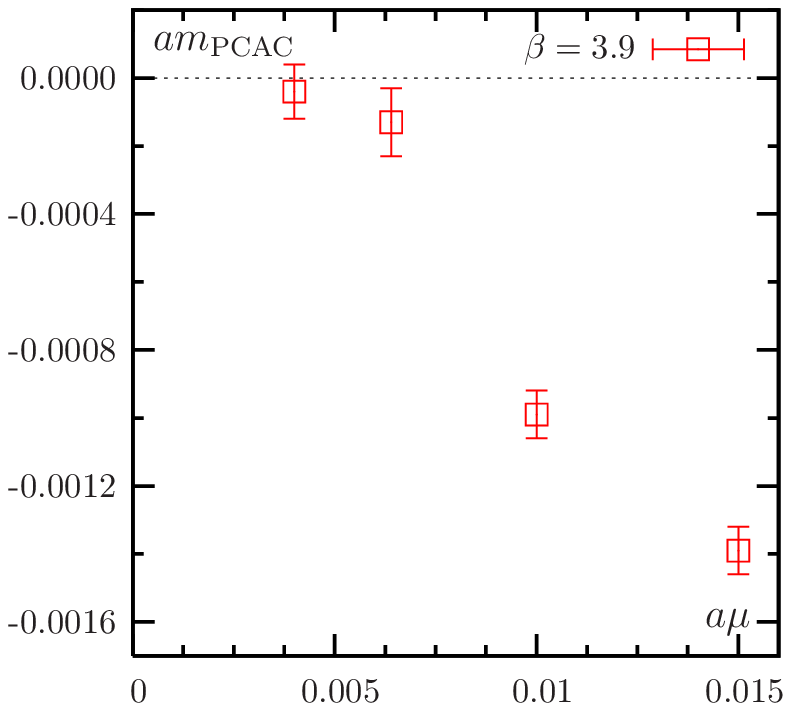}}\quad
	      \subfigure[$r_0$ as a function of $a\mu$ at
		$\beta=3.9$. \label{fig:r0}]
			{\includegraphics[width=0.417\linewidth]{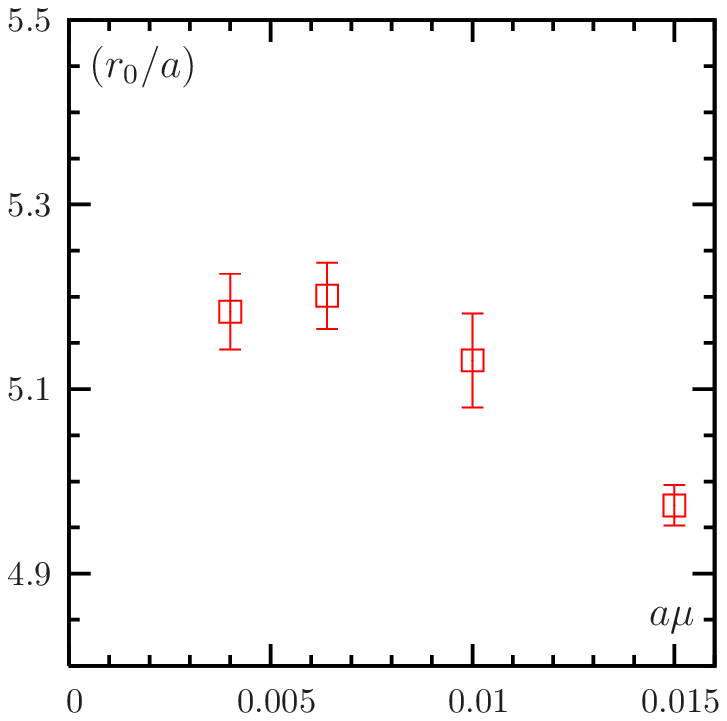}}
			\caption{PCAC quark mass and Sommer parameter as a function of the twisted mass.}
			\label{fig:r0andmpcac}
  \end{figure}

  \subsection{Setting the Scale}

  As mentioned already we used for this first analysis the Sommer
  parameter $r_0$ with a value of $0.5\, \mathrm{fm}$ to set the
  scale. While $r_0/a$ is measurable to high accuracy in lattice QCD
  simulations it has the drawback that its value in physical units is
  not known very well, which might introduce a non-negligible
  uncertainty. Therefore, it could be advantageous to determine the scale
  using other quantities which are experimentally better known, such as 
  $f_\pi$, $f_\mathrm{K}$ or $m_{\mathrm{K}^*}$ which we plan to do in the future.

  In figure \ref{fig:r0} we plot $r_0/a$ as a function of $a\mu$ at
  $\beta=3.9$. Within the current errors the mass dependence of this
  quantity seems to be rather weak for the lowest three values of $a\mu$
  , which would allow for a constant extrapolation to the physical point
  ($m_\pi=137\,\mathrm{MeV}$). 
  Again, for the time being, we use the value of
  $r_0/a$ listed in table \ref{tab:setup} determined 
  at the value of $a\mu_\mathrm{min}$ as an estimate for $r_0/a$ at
  the physical point.

\subsection{$f_\mathrm{PS}$,  $m_\mathrm{PS}$ and other observables}

In this subsection we present first results for $f_\mathrm{PS}$ and
$m_\mathrm{PS}$. The pseudo-scalar mass $m_\mathrm{PS}$ is as usual 
extracted from the
exponential decay of corresponding correlators constructed from local
operators. 
In contrast to pure Wilson fermions, for maximally twisted mass fermions
an exact lattice Ward identity allows to extract the pseudo-scalar decay
constant $f_{\rm PS}$ from the relation
\begin{equation}
  \label{eq:fps}
  f_\mathrm{PS} = \frac{2\mu}{m_\mathrm{PS}^2} |\langle 0 | P^1 (0)
  | \pi\rangle |\, ,
\end{equation}
with no need of computing any renormalization constant since it turns
out that $Z_P = 1/Z_\mu$ \cite{Frezzotti:2000nk}.

  In order to account for finite size (FS) effects in
  $m_\mathrm{PS}$ and $f_\mathrm{PS}$ 
  we used NLO $\chi$PT formulae as derived by
  Gasser and Leutwyler \cite{Gasser:1986vb}. The
  corrections decrease exponentially with a rate $m_\mathrm{PS}\cdot L$ (which we
  take from our data) and are proportional to the overall factor
  $(f_\mathrm{PS}\cdot L)^{-2}$. The quantity $f_\mathrm{PS}\cdot L$ is
  estimated at the chiral point either via extrapolation of uncorrected
  data for $f_\mathrm{PS}$ or from the phenomenological value of 
  $f_\pi$ assuming $r_0 = 0.5\, \mathrm{fm}$. 
  The difference between the results for the
  corrected $f_\mathrm{PS}$ obtained in the two ways at the physical point is
  about $0.4\, \mathrm{MeV}$, which is certainly less than the
  uncertainty arising from neglecting higher orders in $\chi$PT for the 
  FS effects. Using the
  socalled re-summed L\"uscher formulae with 
  NNLO $\chi$PT input (more precisely the data in Tables 3 and 4 of
  Ref.~\cite{Colangelo:2005gd}), we obtained for $af_\mathrm{PS}$
  estimates of the finite size correction factors that are very similar 
  to those coming from simple NLO $\chi$PT, while for $am_\mathrm{PS}$
  the L\"uscher formula leads to somewhat larger corrections. Note that
  for a definite statement about FS corrections, simulations with larger
  lattices would be needed in order to cross-check the $\chi$PT
  predictions. Such simulations are in progress. 

  In figure \ref{fig:mps} we show $(r_0m_\mathrm{PS})^2$ (not FS
  corrected) as a function of the twisted mass $r_0\mu$ for the two
  $\beta$-values available. The dependence is to a very good
  approximation linear in $r_0\mu$ and the data at the two
  $\beta$ values fall approximately on a common line. 
  Note that the renormalization factor $Z_\mu$ has not been taken into account here.
  The dotted line is not a fit to the data, but a line 
  connecting the highest data point in $r_0\mu$ of
  $\beta=3.9$  with the origin and may serve to guide the eye.

  In figure \ref{fig:fps} $af_\mathrm{PS}$ is plotted versus
  $(am_\mathrm{PS})^2$ at $\beta=3.9$ only. In the plot we  give the FS corrected 
  and uncorrected data.
  Clearly, the effect of the FS correction is visible but not huge. 
  In any case, whether we perform a FS correction or not
  the data show a curvature. It is very tempting to perform a fit of the data 
  according to the expectation from $\chi$PT. For this, it would, however, be better 
  to have a simulation at an additional value of $a\mu > a\mu_\mathrm{min}$. 
  Such a simulation is presently in progress and we will report the results in 
  a forthcoming publication \cite{letter:2006}.

  A linear extrapolation of the FS corrected data of $f_\mathrm{PS}$ at
  $\beta=3.9$ towards the physical point ($m_\pi=137\, \mathrm{MeV}$)
  yields a value of $f_\pi=129.0\pm0.5\pm1.0\, \mathrm{MeV}$ (to be
  compared with $f_\pi=131\,\mathrm{MeV}$). The first
  error comes from the errors on $f_\mathrm{PS}$, $m_\mathrm{PS}$ and
  the extrapolation in the quark mass, the second from the uncertainty
  on the value of $r_0/a$. Systematic errors due to lattice artifacts,
  residual finite volume effects and the choice of $r_0 = 0.5\
  \mathrm{fm}$ are not included.

  \begin{figure}[t]
    \centering
    \subfigure[$(r_0m_\mathrm{PS})^2$ as a function of $r_0\mu$ at $\beta=3.9$
      and $\beta=4.05$. \label{fig:mps}]
	      {\includegraphics[width=0.41\linewidth]{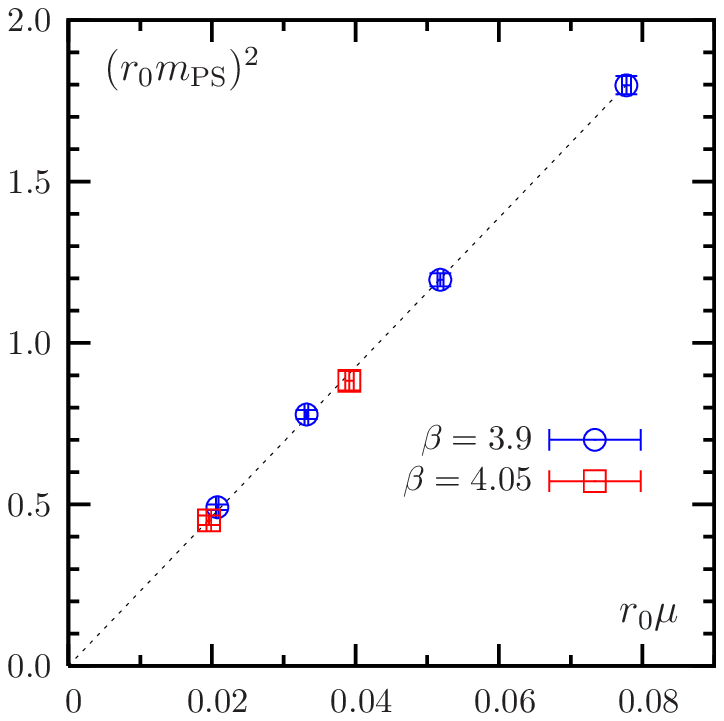}}\quad
	      \subfigure[$af_\mathrm{PS}$ as a function of $(am_\mathrm{PS})^ 2$ at
		$\beta=3.9$ without (squares) and with (circles) FS correction. \label{fig:fps}]
			{\includegraphics[width=0.45\linewidth]{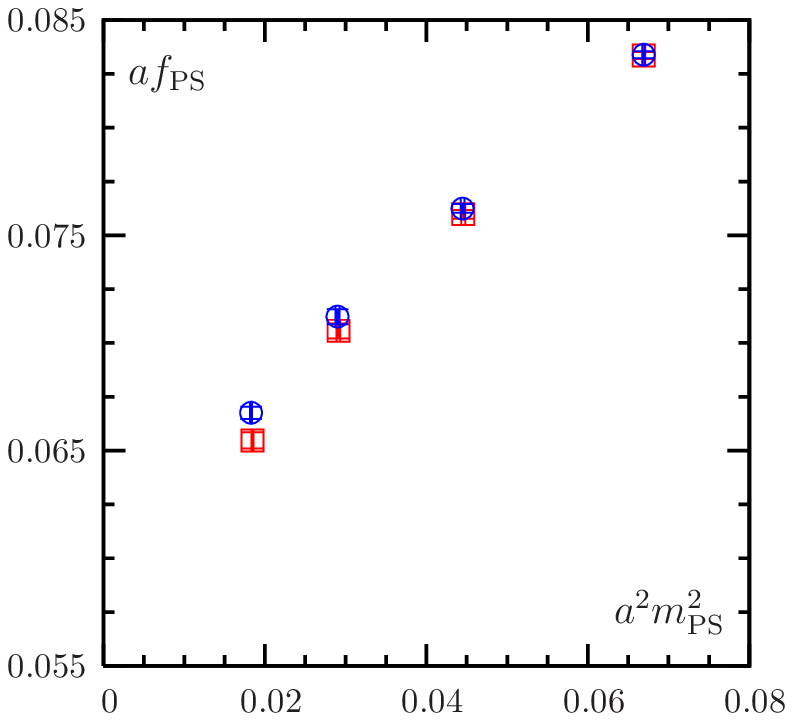}}
			\caption{$(r_0m_\mathrm{PS})^ 2$ as a function of $r_0\mu$ for both
			  $\beta$ values and $af_\mathrm{PS}$ as a function of
			  $(am_\mathrm{PS})^ 2$ for $\beta=3.9$ only.}
			\label{fig:mpsandfps}
  \end{figure}

  In figure \ref{fig:avx} we show a comparison of quenched and dynamical
  values of the average momentum $\langle x\rangle$ corresponding to
  the expectation value of a non-singlet operator in pion states.
  Although the data point from the dynamical simulation still has  
  systematic uncertainties such as missing non-perturbative renormalisation and 
  lack of a continuum extrapolation, the message of the graph is very encouraging: 
  employing twisted mass fermions, it is clearly possible to address the 
  approach to the physical point also for more complicated quantities
  than $m_\mathrm{PS}$ and $f_\mathrm{PS}$ . In addition 
  it seems that such quantities can be computed rather precisely. 

  \begin{figure}[t]
    \centering
    \includegraphics[width=.7\linewidth]{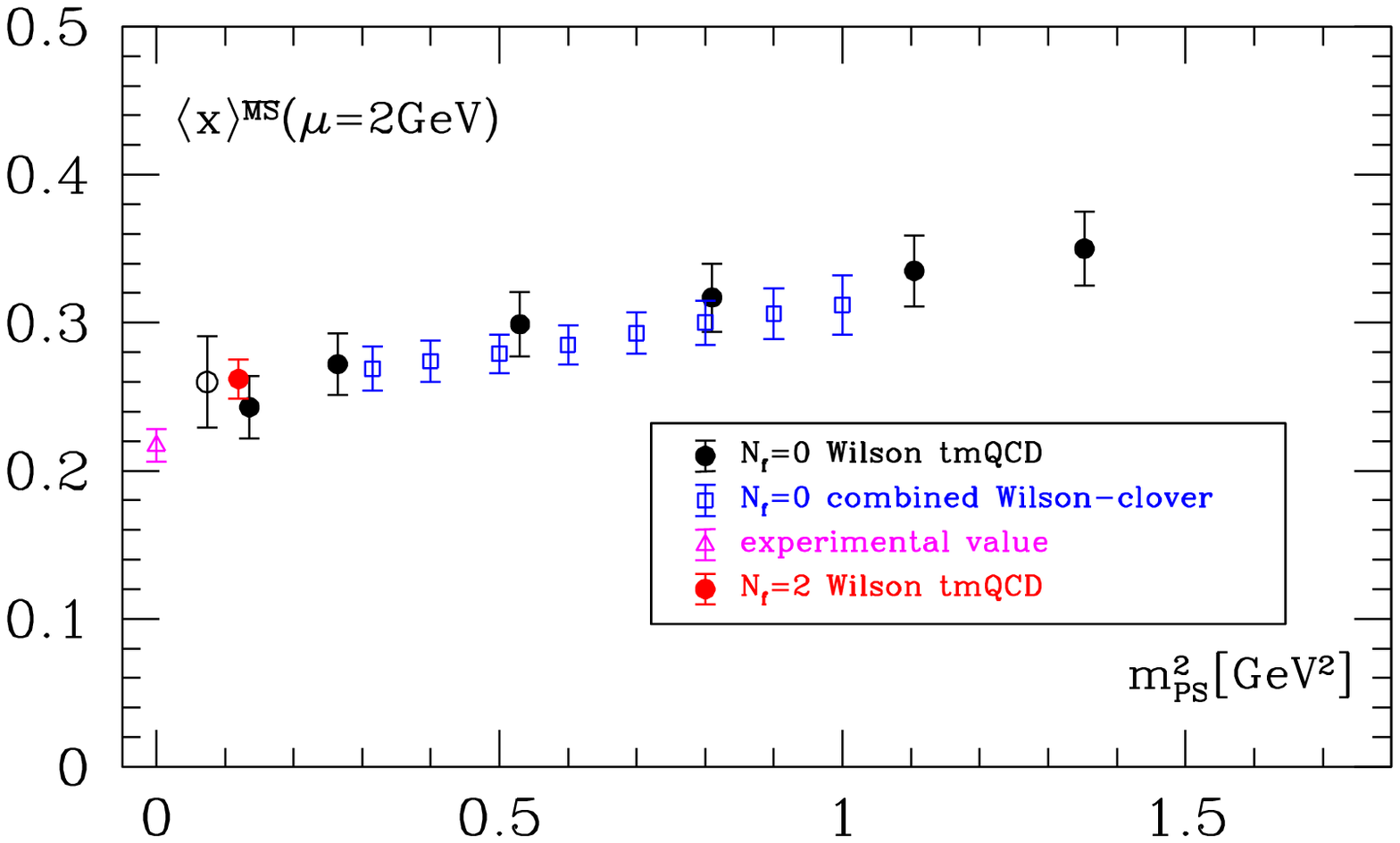}
    \caption{The average momentum of a parton in a pion. The blue squares
      correspond to quenched simulations~\cite{Guagnelli:2004ga} using the
      Schr\"odinger functional. These simulations have controlled non-perturbative
      renormalisation, finite volume effects and continuum limit.
      The black filled circles come from quenched calculations of the same quantity with
      maximally twisted Wilson fermions showing that with this approach
      much smaller quark masses can be reached, from
      Ref.~\cite{Capitani:2005jp} (the open circle corresponds
      to a point where finite size effects may possibly still affect the
      data).
      The red filled circle represents a simulation point obtained {\em with dynamical
	quarks} at $\beta=3.9$ (here the quenched renormalisation factor
      is used). Although a number of systematic effects,
      continuum limit and non-perturbative renormalisation still have to be analysed,
      it is to be noted that also at small quark masses dynamical results
      seem to agree with quenched ones.
      \label{fig:avx}}
  \end{figure}

  \subsection{Effects of Isospin Breaking}

  As mentioned before, a very important issue to address is the effect
  of isospin breaking in 
  the twisted mass formulation at finite lattice spacing. This effect is
  expected to be most significant in the mass splitting between the lightest
  charged and uncharged pseudo-scalar mesons. A first analysis at
  $\beta=3.9$ and $a\mu=0.004$, taking 
  the disconnected contribution in the neutral channel fully into account, 
  shows that the uncharged pseudo-scalar
  meson is about $25\%$ lighter than the charged one. We get, in fact
  \[
  am_\pi^\pm = 0.1361(5)\, ,\qquad am_\pi^0 = 0.103(4)\, ,
  \]
  or, expressed differently, 
  $r_0^2((m_\pi^0)^2-(m_\pi^+)^2)=c(a/r_0)^2$ with $c\approx-5$, which has a
  magnitude that is a factor of 2 smaller than what was found in our
  quenched investigations~\cite{Farchioni:2005hf}. Taking the value of
  $c=-5$ and assuming a linear dependence of the charged and neutral pseudo-scalar 
  mass-squared on $a\mu$, we can compute the value of $a\mu$ where
  the neutral pion becomes massless and the phase transition line
  starts. This critical value would be $a\mu_c\simeq 0.0015$ at $\beta=3.9$ and hence 
  well below the smallest value of $a\mu$ employed in our simulations.
  If we consider the mass splitting in the vector meson sector, we find that 
  this splitting is compatible with zero within our currently still
  large errors.
  Note that the uncharged pion being lighter than the charged one is
  compatible with predictions from $\chi$PT \cite{Scorzato:2004da} in
  case the first order phase transition scenario is realised.

  Let us mention that 
  the disconnected correlations needed for the $\pi^0$ meson are evaluated
  using a stochastic (Gaussian)  volume source with 4 levels of
  hopping-parameter variance reduction \cite{McNeile:2000xx}. We use 24
  stochastic sources per gauge configuration and evaluate every $10$
  trajectories.

  \subsection{Continuum Scaling of $f_\mathrm{PS}$}

  In figure \ref{fig:scaling} we show the
  scaling behaviour of $r_0f_\mathrm{PS}$ with $(a/r_0)^2$ using our 
  current simulation data. 
  These results 
  shown in the plot are still affected by 
  some uncertainties: First of all the values of
  $r_0m_\mathrm{PS}$ are not (exactly) matched because there are only
  two values of $a\mu$ available at $\beta=4.05$. Second, we have taken
  -- as mentioned before -- the value of $r_0/a$ at 
  $a\mu_\mathrm{min}$ at the two $\beta$-values as an estimate of the value 
  at the physical point. 

  However, we do not believe that those uncertainties are larger than
  the current statistical uncertainty. Therefore, figure
  \ref{fig:scaling} suggests that lattice artifacts are rather small in
  $r_0f_\mathrm{PS}$ at the current level of accuracy.
  This result is a first indication that lattice artifacts are small and 
  well under control in simulations with maximally twisted mass
  fermions, although further tests are clearly needed.

  \begin{figure}[t]
    \centering
    \includegraphics[width=.7\linewidth]{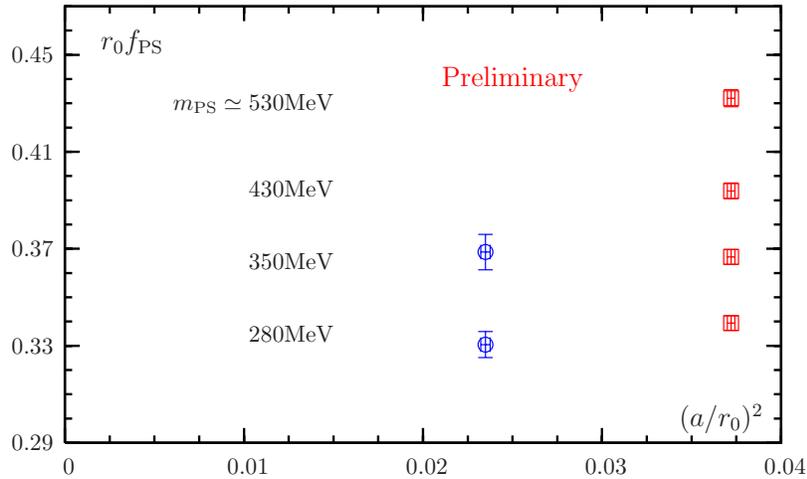}
    \caption{Preliminary scaling plot of $r_0 f_\mathrm{PS}$ with
      $(a/r_0)^2$. We also quote our current estimates of $m_\mathrm{PS}$
      in physical units at $\beta=3.9$.}
    \label{fig:scaling}
  \end{figure}

  \section{Including Strange and Charm Quarks}

  Adding to a doublet of mass-degenerate twisted mass quarks the
  strange quark can be realised in two possible ways.
  Either, a single quark flavour is introduced, or, a non-degenerate
  doublet of quarks is taken, as suggested in Ref.~\cite{Frezzotti:2004wz}.
  We have followed the prescription of Ref.~\cite{Frezzotti:2004wz} where
  it has been shown that a {\em real quark determinant} with
  a mass-split doublet can be obtained if the mass splitting term is taken to
  be orthogonal in isospin space to the twist direction.
  The reason for choosing this solution is that we want to realise 
  $\mathcal{O}(a)$ improvement also for the non-degenerate case by 
  using maximally twisted Wilson fermions. In this way, again 
  no further improvement coefficients have to be computed.

  The action for the non-degenerate quarks can be written in a form
  obtained from the physical basis of Ref.~\cite{Frezzotti:2004wz}
  by means of a chiral rotation around the direction 1 in isospin space
  to undo the twist of the Wilson term. The mass terms become
  \begin{equation}
    S_{m,2+1+1} = a^4\sum_x \big\{\bar{\psi}(x)[\tilde{m}_0 + \mu_\delta\tau^3
      +i\mu_\sigma \gamma_5\tau^1]
    \psi(x)\big\}
    \label{tmQCDnf211}
  \end{equation}
  where $\tilde{m}_0$ has again to be set to its critical value.
  In this case the renormalised current mass of the two quarks
  is given by $(\mu_\sigma/Z_P \pm \mu_\delta/Z_S)$, where
  $Z_P$ and $Z_S$ are the renormalisation constans of the pseudo-scalar
  and scalar non-singlet operators (in the twisted basis).
  The $\psi$-fields couple to the gauge fields through the same Wilson-Dirac 
  operator as the one of Eq.~(\ref{eq:Sf}).

  \begin{figure}[t]
    \centering
    \includegraphics[width=0.7\linewidth]{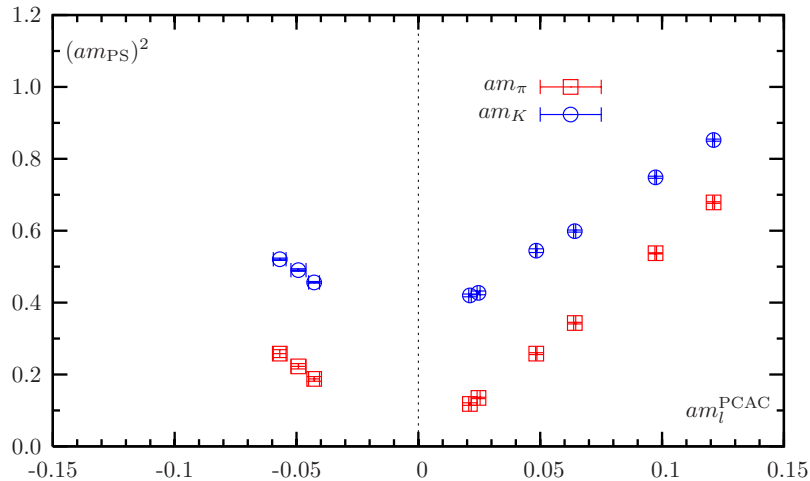}
    \caption{The squared pseudo-scalar pion and kaon masses as a function of
      the PCAC mass $am_l^{PCAC}$ of the light doublet ($a\mu_l=0.0075,
      a\mu_\sigma=0.2363, a\mu_\delta=0.2138$, $16^3\times32$ lattices
      with tlSym gauge action, $\beta=3.35$). Owing to the
      presence of meta-stabilities, 
      there is a gap between positive and negative values of $am_l^{PCAC}$.
      This implies a positive minimal value of the pion mass.}
    \label{massesdyn211}
  \end{figure}

  In our Monte Carlo simulations we described the $u$ and $d$ quarks
  by a \emph{light} mass-degenerate twisted mass doublet and the
  $c$ and $s$ quarks by a \emph{heavy} non-degenerate twisted mass doublet.
  We used the PHMC updating algorithm \cite{Frezzotti:1997ym} 
  with multi-step polynomial
  approximation and stochastic correction in the update chain as
  described in Refs.~\cite{Montvay:2005tj,Scholz:2006hd}.

  As has been discussed in 
  Refs.~\cite{Frezzotti:2003xj,Frezzotti:2004wz,Chiarappa:2006ae}, in
  order to realise $\mathcal{O}(a)$ improvement,
  tuning to full twist can be achieved by setting the values of $\kappa_c$ of
  the light and heavy doublet equal. Therefore
  with respect to the $N_f=2$
  simulations, we are left with only one additional parameter,
  i.e.\ the strange quark mass to be determined by the value of the kaon
  mass, while the charm quark is taken to be much heavier.
  In Ref.~\cite{Chiarappa:2006ae} we give a detailed discussion of
  possible ways to determine the twist angle, renormalisation factors
  and meson masses. 

  We find that                                                        
  simulations of $N_f=2+1+1$ dynamical quarks are feasible
  in practice. The tuning effort is similar to the $N_f=2$ case, modulo
  the need to know the ratio $Z_P/Z_S$ to fix the strange quark mass,
  and turns out to be possible without too much effort.
  Employing the tlSym gauge action as before, we find  
  for lattice spacings as coarse as $a\approx 0.15\,$fm again
  signs of a first order phase transition which appears to be somewhat
  stronger than in the case of $N_f=2$ degenerate quarks.
  As a consequence of this phase transition the
  minimal pion mass cannot be made to vanish. This is illustrated in
  fig.~\ref{massesdyn211}.
  Nevertheless, our simulations allow to estimate the values of $\beta$
  and $\mu$ where the lattice spacing is, say, $a \simeq 0.1\,{\rm fm}$
  and safe simulations with a pion mass below $300\, {\rm MeV}$ become
  possible.
  $\chi$PT formulae for the case of non-degenerate
  quarks have already been developed \cite{Munster:2006yr} and have been confronted to our
simulation results, see Ref.~\cite{Chiarappa:2006ae} for a discussion of this 
point. 

\section{Discussion}

In this proceeding contribution we presented the status of an ongoing
project to simulate two flavour lattice QCD with light maximally
twisted mass quarks. Our first results are very encouraging. 
\begin{itemize}
\item We reach a pseudo-scalar mass of $280\, \mathrm{MeV}$
  (setting the scale with $r_0=0.5\,\mathrm{fm}$) while the simulation is 
  performing smoothly and stable. 
  
\item Comparing results at 2 values of the lattice spacing, we find very good
  scaling, indicating that $\mathcal{O}(a)$ improvement is at work 
  and that also the $\mathcal{O}(a^2)$ lattice artifacts are small. 
  This is in full accordance with earlier results in the quenched approximation.
  
\item We do see effects of isospin breaking which are most significant
  in the mass splitting of the neutral and charged pions and turn out
  to be 25\%. This is already much smaller than the corresponding
  splitting obtained in the quenched approximation.  
  Note that at our smaller value of the lattice spacing, corresponding to 
  $\beta=4.05$, the splitting would be at the 10\% 
  level only, if we assume an $a^2$ dependence of this effect as has been 
  found in the quenched approximation. 

\item We have performed first simulations with non-degenerate strange
  and charm quark flavours in 
  addition to the mass degenerate light doublet. It turns out that tuning and, 
  in particular realising maximal twist, in this 
  case is not problematic and needs a similar effort as other approaches 
  taking the strange quark into account. 
  We have explored the phase structure in this case and 
  determined those parameters where simulations would not be affected by unwanted 
  (first order) phase transitions. Hence, calculations with twisted mass 
  and $N_f=2+1+1$ flavours are well prepared.
\end{itemize}
It has to be said that tuning to full twist has to be performed on the same large
lattice as used later for the calculation of physical 
quantities. The reason for this is that we need to project to the pion state 
as fully as possible to determine the PCAC quark mass without being affected 
by finite size effects or excited state contributions.
Thus the tuning itself is rather expensive. But, it only has to be done once, as 
it is the case for the determination of action improvement coefficients in other 
approaches.
Note, however, that with twisted mass fermions we do not 
have to compute further operator
specific improvement coefficients afterwards. 

Our ongoing efforts concern the completion of the simulations at $\beta=3.9$ 
with an additional value of $a\mu$, which will be presented in a
forthcoming publication \cite{letter:2006}.
The main thrust of the future work is to add another and finer value of the lattice 
spacing (and, maybe, also a coarser one).
In this proceeding contribution we reported already about some first results 
in this direction which indicate only small scaling violations.
Our large set of configurations is uploaded to the ILDG
\cite{webaddress,Jansen:2006ks} and will be used for computations of more 
complicated physical quantities than 
discussed here. Another interesting direction we follow with the twisted mass
approach is to use Osterwalder-Seiler or overlap fermions in the
valence sector \cite{Bar:2006zj} and to perform simulations at
non-vanishing temperature on  which first results were presented at
this conference \cite{mariaposter}.

\subsubsection*{Acknowledgements}
We thank NIC and the computer centres at Forschungszentrum J{\"u}lich
and Zeuthen for providing the necessary technical help and computer
resources. Computer time on UKQCD's QCDOC in Edinburgh
\cite{Boyle:2005fb} using the Chroma code \cite{Edwards:2004sx} and on
apeNEXT \cite{Bodin:2005gg} in Rome and Zeuthen and on MareNostrum in
Barcelona (www.bsc.es) are gratefully acknowledged.
This work has been supported in part by the DFG 
Sonderforschungsbereich/Transregio SFB/TR9-03 
and the EU Integrated Infrastructure Initiative Hadron Physics (I3HP)
under contract RII3-CT-2004-506078. 
We also thank the DEISA Consortium (co-funded by the EU, FP6 project 508830),
for support within the DEISA Extreme Computing Initiative (www.deisa.org).

\bibliographystyle{JHEP-2}
\bibliography{bibliography}

\end{document}